\documentclass{emulateapj}

\newcommand{\myemail}{ymiki@ccs.tsukuba.ac.jp}

\slugcomment{last updated on 2014/01/16(Thu) 09:52:13
}

\shorttitle{Hermitage of Wandering Black Hole in the M31 Halo}
\shortauthors{Miki et al.}

\begin{document}


\title{Hunting A Wandering Supermassive Black Hole in M31 Halo -- Hermitage of Black Hole}


\author{Yohei Miki$^{1}$, Masao Mori$^{1}$, Toshihiro Kawaguchi$^{2}$, and Yuriko Saito$^{3, 4}$}
\affil{$^1$Center for Computational Sciences, University of Tsukuba, 
  Tsukuba, Ibaraki 305-8577, Japan; 
  \myemail}
\affil{$^2$Dept. of Physics and Information Science, Yamaguchi University, 
  1677-1 Yoshida, Yamaguchi, Yamaguchi 753-8512, Japan}
\affil{$^3$Department of Astronomical Science, The Graduate University for Advanced Studies (SOKENDAI), 
  Osawa, Mitaka, Tokyo 181-8588, Japan}
\affil{$^4$National Astronomical Observatory of Japan, 
  Osawa, Mitaka, Tokyo 181-8588, Japan}


\begin{abstract}
In the hierarchical structure formation scenario, galaxies enlarge through multiple merging events with less massive galaxies. 
In addition, the Magorrian relation indicates that almost all galaxies are occupied by a central supermassive black hole (SMBH) of mass $10^{-3}$ of its spheroidal component. 
Consequently, SMBHs are expected to wander in the halos of their host galaxies following a galaxy collision, although evidence of this activity is currently lacking. 
We investigate a current plausible location of an SMBH wandering in the halo of the Andromeda galaxy (M31). 
According to theoretical studies of $N$-body simulations, some of the many substructures in the M31 halo are remnants of a minor merger occurring about 1 Gyr ago. 
First, to evaluate the possible parameter space of the infalling orbit of the progenitor, we perform numerous parameter studies using a Graphics Processing Unit (GPU) cluster. 
To reduce uncertainties in the predicted position of the expected SMBH, we then calculate the time evolution of the SMBH in the progenitor dwarf galaxy from $N$-body simulations using the plausible parameter sets. 
Our results show that the SMBH lies within the halo ($\sim$20--50 kpc from the M31 center), closer to the Milky Way than the M31 disk. 
Furthermore, the predicted current positions of the SMBH were restricted to an observational field of $0\degr.6 \times 0\degr.7$ in the northeast region of the M31 halo. 
We also discuss the origin of the infalling orbit of the satellite galaxy and its relationships with the recently discovered vast thin disk plane of satellite galaxies around M31. 
\end{abstract}


\section{Introduction}
\label{section:introduction}
In a cold dark matter (CDM) universe, the hierarchical structure formation scenario posits that large galaxies, such as the Milky Way and the Andromeda galaxy (M31), 
have enlarged through multiple mergers with smaller galaxies. 
Furthermore, the mass of the spheroidal component of galaxies is correlated with the mass of their central supermassive black holes (SMBHs). 
The Magorrian relation \citep{Magorrian1998, MarconiHunt2003} implies that galaxies coevolve with their central SMBHs. 
However, the coevolution process of galaxies and SMBHs is largely unknown. 
In the hierarchical structure formation scenario, galaxies collide and merge with other galaxies and subsequently with less massive galaxies, causing their central SMBHs to drift within the halo region of their host galaxy. 
In other words, SMBHs wander in the halo of their host galaxy after galaxy merging events and finally sink toward the central region of the host galaxy under dynamical friction. 
Therefore, SMBHs can be either centralized in their host galaxy, as in active galactic nuclei, or reside outside of the nucleus \citep{Bellovary2010}, although evidence of the latter class of SMBHs is currently lacking. 
The search for wandering SMBH has recently attracted great interest \citep{Farrell2009, Wiersema2010}. 
In this study, we theoretically investigate the probable positions of such SMBHs. 

Cosmological $N$-body simulations of the hierarchical structure formation have revealed a wealth of merger remnants around host galaxies \citep[e.g.][]{BullockJohnston2005}. 
To verify theoretical predictions from the CDM scenario, and therefore test the current cosmology, many observational researchers have focused on merger remnants \citep[e.g.][]{Chiba2005, Minezaki2009}. 
A giant stellar stream has been discovered in the southern region of the halo of M31 \citep{Ibata2001}. 
Further photometric and spectroscopic observations of the spatial and radial velocity distributions of red giant stars, as well as the metallicity distribution, have revealed other substructures near M31 \citep{Ferguson2002, McConnachie2003, Ibata2004, Ibata2005, Irwin2005, Guhathakurta2006, Kalirai2006a, Kalirai2006b, Gilbert2007, Ibata2007, Koch2008, Gilbert2009, McConnachie2009, Kalirai2010, Tanaka2010, Richardson2011, Tollerud2012, Martin2013}. 
The motions of test particles under the gravitational potential of M31 have been calculated \citep{Ibata2004, Font2006}, and the interaction between the progenitor of the stream and M31 has been investigated in $N$-body simulations \citep{Fardal2006, Fardal2007, Fardal2012, Fardal2013, MoriRich2008, Sadoun2013, Miki2013?}. 
These studies suggest that the stream, the northeast shell, and the west shell constitute tidal debris formed during recent pericentric passages of a radially-accreting satellite. 
These models reproduce the observed features and successfully restrain the orbit and the properties of the progenitor. 

From the Magorrian relation, a progenitor dwarf galaxy is expected to be occupied by an SMBH of mass $M_{\rm BH}$ about $10^{-3}$ times the mass of the spheroidal component of its host galaxy $M_{\rm sph}$ \citep{Magorrian1998, MarconiHunt2003}. 
A similar relation between $M_{\rm BH}$ and the velocity dispersion $\sigma$ of the host galaxy, called the $M_{\rm BH}-\sigma$ relation, has been confirmed down to $M_{\rm BH} \sim 10^5 M_\odot$ \citep{Barth2005, Xiao2011}. 
Therefore, the relation between SMBHs and their host galaxies holds down to $M_{\rm sph} \sim 10^8 M_\odot$. 
Since the dynamical mass of the progenitor is estimated to be of order $10^9 M_\odot$ \citep{Fardal2007, Fardal2013, MoriRich2008, Miki2013?}, the progenitor likely harbored an SMBH of mass up to $10^6 M_\odot$; assuming that the progenitor consisted solely of a spheroidal stellar component. 
Thus, an SMBH should currently be wandering among the merger remnants. Finding such an SMBH will help to elucidate how SMBHs coevolve with galaxies. 
This study uses $N$-body simulations to predict the current likely positions of the SMBH and thereby guide future observational detections. 

Using $N$-body simulations, we first constrain the infalling orbit of the progenitor satellite galaxy (\S\ref{section:lrs}). 
In \S\ref{section:hrs}, we investigate the current plausible position of the expected SMBH in the M31 halo and derive an observational field for future observations. 
The physical relationships between this merger event and the satellite galaxy distribution around M31 is discussed in \S\ref{section:discussion}. 
The study is summarized in \S\ref{section:conclusion}. 

\section{Infalling Orbit of the Satellite}
\label{section:lrs}
In this section, low-resolution $N$-body simulations are conducted over a wide parameter range to restrict the infalling orbit of the progenitor satellite galaxy. 
Following this investigation, $N$-body simulations are conducted at higher resolution in \S\ref{section:hrs}. 
The numerical modeling and analysis are discussed in \S\ref{subsec:lrs:modeling} and \S\ref{subsec:lrs:analysis}, and simulation results are presented in \S\ref{subsec:lrs:result}. 

Before detailing our numerical modeling technique, we emphasize how it differs from that of an earlier parameter study focused on the infalling orbit of the satellite \citep{Fardal2013}. 
\citet{Fardal2013} sought the parameter set that best reproduced the observed structures. 
They were interested in the physical properties of M31 and its progenitor satellite. 
Consequently, they focused on a very narrow region of parameter space around the best-fit configuration. 
By contrast, we seek to restrict the region in which an SMBH of unknown location wanders around the M31 halo. 
This requires a wider and more systematic exploration of the parameter phase space to identify all plausible parameter ranges. 
Our systematic parameter study over a wide parameter region complements the study of \citet{Fardal2013}. 

\subsection{Numerical Modeling of M31 and the Satellite}
\label{subsec:lrs:modeling}
We simulate an accreting satellite dwarf galaxy interacting with M31 using $N$-body simulations, concentrating on the infalling orbit of the progenitor dwarf galaxy. 
We assume an axisymmetric fixed potential model composed of a Hernquist bulge \citep{Hernquist1990}, an exponential disk, and an NFW halo \citep{NFW1996} for M31 \citep{Geehan2006, Fardal2007}. 
This assumption of the fixed potential model is appropriate because the dynamical response of M31's disk to progenitor collision is negligible if the dynamical mass of the progenitor is below $5\times 10^9 M_\odot$ \citep{MoriRich2008}. 
Our numerical simulations are performed in a Cartesian coordinate system $(x, y, z)$ whose origin represents the center of M31. 
The $z$ axis is directed along our line-of-sight, and the $x$ and $y$ axes point east and north on the sky plane, respectively. 
This coordinate system has been commonly adopted in earlier studies of M31. 
The distance from Earth to M31 is assumed as 780 kpc \citep{McConnachie2003}; thus, $1\degr$ corresponds to a physical scale of 13.6 kpc. 
The heliocentric velocity of M31 toward the line-of-sight, east and south on the sky plane is assumed as $-300$ km s$^{-1}$ \citep{deVaucouleurs1991}, $127$ km s$^{-1}$ and $75$ km s$^{-1}$ \citep{Sohn2012, vanderMarel2012}, respectively. 

By restricting the area in which the SMBH exists, this study aims to determine the observational field for future observational detections. 
The greatest contributor to the uncertainty in the current SMBH position is uncertainty of the infalling orbit of the progenitor dwarf galaxy. 
Therefore, we should perform a large parameter study in the six-dimensional phase space to constrain the orbit of the infalling satellite to that of the observed structures. 
Since a six-dimensional parameter space is excessively large for an exhaustive search, even by recent high-performance computer architectures, we reduce the number of dimensions as follows. 
First, to ensure that the satellite interacts with M31, we fix the initial distance of the infalling satellite as $7.63$ kpc from the center of M31 \citep[corresponding to the scale radius of a dark matter halo;][]{Fardal2007}. 
In addition, we model M31 as an axisymmetric system. 
Imposing these conditions, the parameter space is reduced to four dimensions (the altitude of the initial position and the initial velocity vector), but remains very large. 

To realize a sufficiently wide parameter space, the parameter sets are distributed on a relatively coarse grid defined in M31-centric spherical coordinates. 
Since M31 is axisymmetric, the azimuthal angle (around the rotation axis of the M31 disk) of the initial satellite position is simply related to the observational angle. 
Instead of performing multiple $N$-body simulations at different azimuthal angles, we ``observe'' snapshots of $N$-body simulations by rotating around the axis with a $\sim$3$\degr$ bin width. 
To determine the altitude of the initial position, the northern hemisphere of M31 is covered in $6\degr$ increments. 
We focus on a merger occurring immediately prior to the giant stellar stream in the southern hemisphere; thus, we consider the initial orbital position to lie in the northern hemisphere. 
The possibility that the satellite entered from the opposite side is discussed in \S\ref{section:discussion}. 

The grid width of the infalling velocity of the satellite in the radial direction is $\sim$13 km s$^{-1}$. 
Two characteristic velocities are $550$ and $440$ km s$^{-1}$, respectively, specifying the escape velocity at the initial position ($7.63$ kpc from the center of M31), and the velocity required to set the apoapsis of the satellite at $r=100$ kpc. 
Since the giant stellar stream extends beyond 100 kpc from the center of M31 \citep{McConnachie2003}, the infalling velocity of the satellite should exceed $440$ km s$^{-1}$. 
For a fair evaluation of other possibilities, we also test models in which the infalling radial velocity exceeds $550$ km s$^{-1}$ and is slower than $440$ km s$^{-1}$. 
The tangential velocity vector at the initial position is described by two parameters specifying its norm and direction. 
The former parameter (the speed of the tangential velocity) is varied as a ratio of the radial velocity from 0 to 0.47 in $\sim 6.7 \times 10^{-2}$ increments. 
The latter parameter (direction of the tangential velocity) covers $360\degr$ divided into 32 bins. 

In this parameter survey, the infalling satellite contains $65,536$ particles and the gravitational softening length is set as 50 pc. 
The progenitor is assumed as a King sphere with total mass $M_{\rm tot}=3\times 10^9 M_\odot$, concentration $c = 0.7$, and tidal radius $r_{\rm t} = 4.5$ kpc, since this model best matches the progenitor dwarf galaxy when the progenitor follows Fardal's orbit \citep{Miki2013?}. 
The numerical scheme adopts a second-order leapfrog integrator and a shared fixed time step.

To sweep such a wide parameter space, we perform a vast parameter survey utilizing a Graphics Processing Unit (GPU) cluster, namely, HA-PACS at the University of Tsukuba. 
For this purpose, a highly optimized $N$-body simulation code is implemented on the GPU cluster \citep{Miki2013}. 
The combination of the state-of-the-art architecture and the highly optimized code enable the parameter study. 
HA-PACS, which equips over thousand boards of NVIDIA Tesla M2090, is a desirable system to sweep the wide parameter space. 
Furthermore, the code has a peak performance of 1 TFlop/s in single precision with a single NVIDIA Tesla M2090 board, which is 76\% of the theoretical peak performance. 
Around thousand runs of low-resolution $N$-body simulations complete in a day when 128 boards of NVIDIA Tesla M2090 (about one-eighth of HA-PACS) are in use.

\subsection{On-the-fly Analysis}
\label{subsec:lrs:analysis}
Since the number of $N$-body simulation runs is prohibitively large, we automatically and simultaneously analyze each snapshot of the numerical simulations. 
A check list of the automatic online evaluation is provided below:

\begin{enumerate}
\item
  The stellar stream and the west shell exist, and each mass exceeds $10^7 M_\odot$. 
  This minimum value is much smaller than $2.4 \times 10^8 M_\odot$ estimated by \citet{Fardal2006}, who assumed $M/L_{\rm V} \approx 7$.
\item
  The stellar stream is the most luminous structure in the southern area. 
  The giant stellar stream is the most luminous object found by the PAndAS project in this region \citep{McConnachie2009,Martin2013}. 
  This criterion eliminates the event of the undiscovered former satellite surviving the collision with M31.
\item
  The position of the surface density peak of the stellar stream matches that of the observed peak: the density of the simulated stream must peak within a fan-like region of angular width $15\degr$, containing the observation field of the giant stellar stream \citep{Font2006}. 
  This condition is similar to that described in \citet{Fardal2013}.
\item
  The shapes of the two stellar shells adequately agree with the observed shapes. 
  To quantify how precisely each run reproduces the observed shapes of the two shells, we compute the reduced $\chi^2$ given by
  \begin{equation}
    \chi_\nu^2 \equiv \frac{1}{\nu} \sum_{i = 0}^{N-1} \left(\frac{x_{i,\, {\rm sim}}-x_{i,\, {\rm obs}}}{\sigma_{i,\, {\rm obs}}}\right)^2,
    \quad
    \nu = N-1,
    \label{eq:def:reduced.chi.square}
  \end{equation}
  where the number of sampling points $N$ is $48$. 
  The edge position of the observed shells is evaluated from the star count map prepared by \citet{Irwin2005}. 
  Successful parameter sets must satisfy $\chi_\nu^2 \leq 1.7$ \citep[$99.7$\% confidence level according to][]{Press2007}. 
\item
  The sharpness of the edge of the two stellar shells is consistent with the observations. 
  We stipulate that the stellar density inside the edge is more than two times the stellar density outside the edge. 
\item
  The mass-density ratios among the stellar stream, the east shells and the west shells are similar to the observed ratios. 
  Subtracting the noise from the observed star count map by \citet{Irwin2005}, we obtain the number density ratios of the east shell over the stream and the east over the west shells as $1.77 \pm 1.57$ and $2.05 \pm 1.80$, respectively. 
  We stipulate that the mass density ratio is within $1\sigma$ scatter. 
\end{enumerate}

To ``observe'' the simulated snapshots, we must assume that the mass-to-light ratio of the satellite galaxy is evaluated in the visible light range ($V$-band); i.e., $M_{\rm tot}/L_{\rm V}$ where $L_{\rm V}$ is $V$-band luminosity. 
Based on the Faber-Jackson relation in the nearby universe \citep{FalconBarroso2011, Toloba2012}, the estimated $V$-band magnitude of the satellite galaxy is $-17.73 \pm 0.69$ (corresponding to $M_{\rm tot}/L_{\rm V}$ of $2.84^{+2.52}_{-1.34}$). 
To eliminate the effects of large uncertainty in the mass-to-light ratio, we assume the bright-end of the Faber-Jackson relation ($M_{\rm tot}/L_{\rm V}=1.51$) and ``observe'' as many faint structures as possible. 
To mimic the observed star count map \citep{Irwin2005}, we ``observe'' the numerical results imposing a limiting magnitude of $V=24. 5$ \citep[the detection limit of the Wide Field Camera on the Isaac Newton Telescope;][]{Irwin2005}. 

The method for detecting the edge of the stellar shell while ``observing'' snapshots is optimized to capture all edge-like features. 
All density peaks and valleys along a radial direction on the sky plane are assigned as edge candidates, and the candidate nearest to the observed edge is tagged as the ``observed'' edge in the snapshots. 
If we know the actual mass-to-light ratio of the progenitor satellite galaxy, then the easiest and most plausible way to determine the shell edges is to combine the mass-to-light ratio with the instrument detection limit.

This simple method, however, would miss some edge signatures if the mass-to-light ratio is assumed greater (i.e. the satellite is assumed fainter) than the actual ratio. 
To avoid this situation, we detect the edge of the shells by the abovementioned method. 
Since only a small number of the $N$-body particles are used, spurious density peaks and valleys are introduced by Poisson noise, which artificially decreases the reduced $\chi^2$ value. 
Later, this effect will be eliminated in the high-resolution $N$-body simulations (see \S\ref{section:hrs}). 

Here, we compare the above ``observing'' criteria with those of earlier studies. In all of the earlier studies, the structures formed after a galactic merger had reproduced (all or some of) the observed global shapes [minor merger scenarios by \citet{Fardal2007,Fardal2012,Fardal2013,MoriRich2008,Sadoun2013,Miki2013?}, and major merger scenario by \citet{Hammer2010,Hammer2013}]. 
Most studies have compared the shapes of the simulated and observed structures by human eyes; exceptions are \citet{Fardal2013} and our previous study \citet{Miki2013?}. 

\subsection{Constraints on the Orbit of the Satellite}
\label{subsec:lrs:result}
\begin{figure}
  \centering
  \epsscale{1.18}
  \plotone{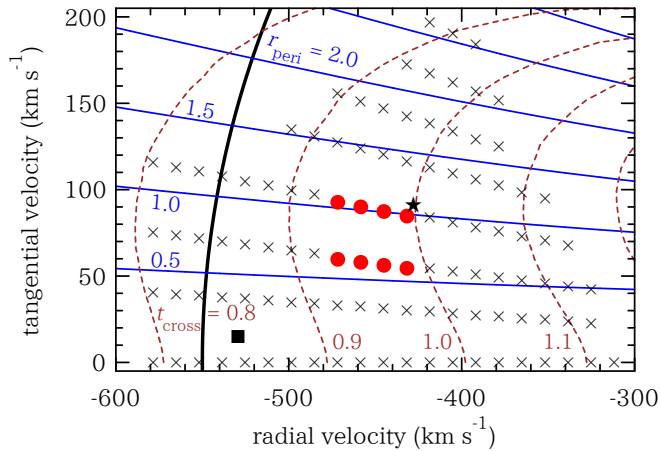}
  \caption{
    Results of $44,880$ runs of the low-resolution parameter study (corresponding to $\sim$5.7 million orbit models), simulating the infalling orbit of the satellite. 
    The horizontal and vertical axes are the infalling radial and tangential velocities, respectively, of the satellite 7.63 kpc distant from the center of M31. 
    Red filled circles indicate that the results accurately reproduce the observed structures. 
    Crosses represent results that failed to reproduce the observed structures. 
    Overlaid curves show contour maps of periapsis, $r_{\rm peri}$ (kpc; blue solid lines), and crossing time $t_{\rm cross}$ (in units of free-fall time of the satellite; brown dashed lines). 
    The filled star and the filled square represent the infalling orbit of the progenitor satellite galaxy in earlier studies \citet{Fardal2007} and \citet{Sadoun2013}, respectively. 
    Thick solid curve corresponds to the escape velocity $550$ km s$^{-1}$ at $7.63$ kpc away from the center of M31. 
  }
  \label{fig:survey.result}
\end{figure}

We performed $44,880$ runs of $N$-body simulations, corresponding to $5,699,760$ models of the infalling orbit of the progenitor satellite. 
The results of the $5,699,760$ orbit models are shown in Fig.~\ref{fig:survey.result}. 
By automatic ``observation'' and by analyzing the $N$-body simulations described in \S\ref{subsec:lrs:analysis}, we identified 138 orbit models that accurately reproduced the observed structures, hereafter referred to as orbit candidates (filled circles in Fig.~\ref{fig:survey.result}). 
More specifically, only 138 out of $5,699,760$ infalling orbit models passed our tests in this low-resolution parameter study. 
In other words, the possible parameter space is an extremely narrow region of the phase space. 
Periapsis $r_{\rm peri}$ (contour map of solid curves) and $t_{\rm cross}$ (contour map of dashed curves) were evaluated by test-particle calculation under a fixed potential in the spherical components of M31 (bulge and halo). 
The crossing time is defined as the time required to pass the region $r \leq r_{\rm c} = 4.3\, {\rm kpc}$ at the first pericentric passage, where $r_{\rm c}$ is the critical radius from the center of M31. 
At $r_{\rm c}$, the core radius of the satellite corresponds to its Hill radius against the M31 bulge. 
Within this radius, the tidal force exerted by M31 largely governs the time evolution of the satellite. 
At the core radius of the satellite, the free-fall time $t_{\rm ff}$ is 15 Myr. 

The distribution of orbit candidates in Fig.~\ref{fig:survey.result} is concentrated around $r_{\rm peri} \cong 0.6-1$ kpc and $t_{\rm cross} \cong 0.95 ~t_{\rm ff}$. 
Tidal forces exerted by the bulge of M31 stretch and disrupt the infalling satellite. 
Since the strength of the tidal force from the bulge of M31 is proportional to $r^3_{\rm peri}$, the parameters that adequately match the observed structures are restricted to a narrow $r_{\rm peri}$ region. 
Figure~\ref{fig:survey.result} shows that $t_{\rm cross}$ is also tightly constrained, implying the importance of the satellite's dynamical response to the tidal force exerted by the bulge of M31. 
The strong tidal field in the bulge of M31 ensures that even a small difference of the crossing time markedly affects the present-day structures. 

One of the most important results in this study is that a maximum infall velocity of the progenitor satellite galaxy exists ($\sim$480 km s$^{-1}$). 
Since the escape velocity is 550 km s$^{-1}$ (thick solid curve in Fig.~\ref{fig:survey.result}), the observed structures can only be formed by an M31-bound satellite galaxy. 
The collision that occurred several hundred megayears ago should have been the first collision of the infalling satellite, because the strong tidal field exerted by the bulge of M31 will destroy the satellite in a single passage. 
However, as noted by \citet{Sadoun2013}, this situation does not naturally arise in the hierarchical CDM context. 
This controversy and its solution will be described in \S\ref{section:discussion}. 

We now compare our results with those of related studies \citep{Fardal2007, Sadoun2013}. 
The infalling orbit found by \citet{Fardal2007} (star in Fig.~\ref{fig:survey.result}) locates near the edge of the area occupied by the 138 orbit candidates. 
This indicates consistency between our study and that of \citet{Fardal2007}. 
Contrarily, the infalling orbit of \citet{Sadoun2013} locates outside of our area. 
\citet{Sadoun2013} set the satellite distant from M31, to delay its collision with M31. 
This discrepancy between our study and \citet{Sadoun2013} chiefly arises from the strict criteria adopted in our study, especially the reduced $\chi^2$ analysis imposed on the shapes of the observed two stellar shells.

\section{Infalling Orbit of the SMBH}
\label{section:hrs}
Here, we discuss high-resolution $N$-body simulations of the 138 orbit candidates that survived the low-resolution parameter study described in the previous section. 
First, we explain the numerical differences between the high- and low-resolution models in \S\ref{subsec:hrs:modeling}. 
The resulting positions of the SMBH derived from the high-resolution $N$-body simulations are presented in \S\ref{subsec:hrs:hermitage} and \S\ref{subsec:hrs:locus}. 

\subsection{Numerical Modeling with the SMBH}
\label{subsec:hrs:modeling}
The purpose of this section is to restrict the locality of the wandering SMBH in the host galaxy. 
To simultaneously reproduce the observed structures and track the orbit of the SMBH, we perform high-resolution $N$-body simulations of M31 interacting with a progenitor satellite containing an SMBH. 
We set the number of particles in the satellite to $524,288$, and the gravitational softening length to 13 pc (equivalent to $\sim$10 increase in the number of particles with $\sim$1/4 softening length, relative to the low-resolution survey). 
Here, the SMBH is represented by an additional particle of mass $3\times10^6 M_\odot$ ($\sim$500 times more massive than the other $N$-body particles), which is placed at the center of the progenitor dwarf galaxy. 
This mass derivation assumes that the progenitor's stellar mass corresponds to its dynamical mass and that the Magorrian relation ($M_{\rm BH} \sim 10^{-3} M_{\rm sph}$) holds. 
Specifically, we adopt the maximum mass of the SMBH. 
All the other parameter setups are the same with that for the low-resolution $N$-body simulations, and as well the computation performs on HA-PACS. 

\subsection{Hermitage of the SMBH}
\label{subsec:hrs:hermitage}
\begin{figure}
  \centering
  \epsscale{1.18}
  \plotone{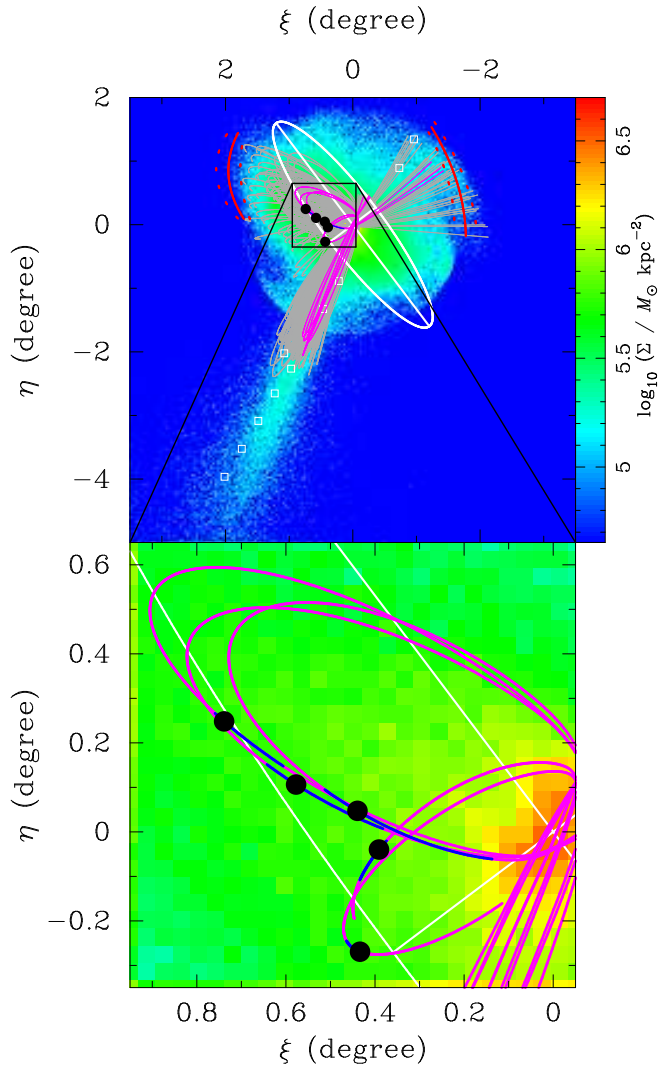}
  \caption{
    Mass distribution (column density) maps of the debris of the dwarf galaxy in standard coordinates centered on M31. 
    The color scale is shown along the right vertical axis of the upper panel. 
    In the upper panel, global distribution of $N$-body particles (the best-fit epoch of the orbit model corresponding to SMBH ID 3 in Tab.~\ref{tab:mbh}) is shown as a color image while white curves and lines show the M31 disk. 
    White squares show observed points of the giant stellar stream \citet{Font2006}, and red curves show the edge and the width of the observed shells. 
    The lower panel is a $1\degr \times 1\degr$ enlarged view of the black square in the upper panel. 
    Black circles show the most probable current position of the SMBH. Gray and magenta curves show the orbits of the SMBH particles for the 138 orbit candidates and the 5 successful candidates, respectively. 
    Blue curves show the SMBH positions when the observed shells are reproduced at the $99.7\%$ confidence level. 
    }
  \label{fig:maps}
\end{figure}
\begin{deluxetable*}{ccccccccc}
\tabletypesize{\scriptsize}
\tablewidth{0pt}
\tablecaption{Summarized information of five SMBH particles at the best-fit epoch \label{tab:mbh}}
\tablehead{ 
  \colhead{ID} 
& \colhead{$\chi_\nu^2$\tablenotemark{(a)}} 
& \colhead{$\xi$ (degree)\tablenotemark{(b)}} 
& \colhead{$\eta$ (degree)\tablenotemark{(b)}} 
& \colhead{$r_{\rm M31}$ (kpc)\tablenotemark{(c)}} 
& \colhead{R.A. (J2000.0)} 
& \colhead{Decl. (J2000.0)}
& \colhead{$D$ (kpc)\tablenotemark{(d)}} 
& \colhead{$v_{\rm los}$ ($\rm{km\, s}^{-1}$)\tablenotemark{(e)}} 
}
\startdata
1 & $1.42$ & $0.74$ & $ 0.25$ & $27.3$ & $00~45~41.60$ & $+41~31~02.16$ & $754.8$ & $-131.0$\\
2 & $1.51$ & $0.39$ & $-0.04$ & $48.9$ & $00~44~18.16$ & $+41~13~45.71$ & $731.4$ & $-355.4$\\
3 & $1.56$ & $0.43$ & $-0.27$ & $36.8$ & $00~44~28.43$ & $+40~59~59.54$ & $743.8$ & $-178.1$\\
4 & $1.59$ & $0.58$ & $ 0.11$ & $22.0$ & $00~45~02.82$ & $+41~22~31.47$ & $759.4$ & $- 77.7$\\
5 & $1.61$ & $0.44$ & $ 0.05$ & $17.9$ & $00~44~29.79$ & $+41~18~58.52$ & $763.2$ & $- 33.1$
\enddata
\tablenotetext{(a)}{Reduced $\chi^2$ when matching observed and simulated shapes of the two stellar shells.}
\tablenotetext{(b)}{Position in M31-standard coordinates. }
\tablenotetext{(c)}{Distance from the center of M31. }
\tablenotetext{(d)}{Distance from the Local Standard of Rest. }
\tablenotetext{(e)}{Heliocentric line-of-sight velocity. }
\end{deluxetable*}

In the high-resolution $N$-body simulations, five orbit models out of the 138 orbit candidates accurately reproduced the observed structures. 
The increased number of particles yields a smoother mass distribution than in the low-resolution calculations, since Poisson noise is reduced. 
This effect eliminates orbit models whose reduced $\chi^2$ values were underestimated when matching the shapes of the observed shells, as discussed in \S\ref{subsec:lrs:analysis}. 
Figure~\ref{fig:maps} shows the mass distribution map of the debris of the satellite galaxy, obtained by the best-fit epoch of the orbit model corresponding to SMBH ID 3 in Tab.~\ref{tab:mbh}. 
As shown in the upper panel of Fig.~\ref{fig:maps}, $N$-body simulations accurately reproduce the observed structures, and the SMBH exerts no significant effect on the formation of the global structures. 
The top panel of Fig.~\ref{fig:maps} is overlaid with the orbits of the SMBH particles for the 138 orbit candidates (gray curves on the mass distribution map). 
Magenta curves show the orbits of the SMBH particles for the five successful candidates. 
The five black circles show the most probable current positions of the SMBH in the corresponding orbit models. 
Positional and velocity information of the five SMBH particles is summarized in Table~\ref{tab:mbh}. 
The candidates are listed in ascending order of reduced $\chi^2$ at the best-fit epoch. 

The lower panel of Fig.~\ref{fig:maps} is an enlargement of the black hatched region of the upper panel, covering a region of $1\degr \times 1\degr$ ($\sim$15 kpc $\times$15 kpc). 
The blue curves trace the orbits of the SMBH particles when the observed shells are reproduced at the 99.7\% confidence level. 
These curves indicate the possible regions currently occupied by the SMBH. 
Clearly, the blue curves are confined to a small region ($\sim 0\degr.6 \times 0\degr.7$), so the candidate field in which the SMBH must exist is within $1\degr \times 1\degr$. 
The above tight constraint for the current position of the wandering SMBH is imposed by strong constraints on the following two factors: 1. the infalling orbit of the progenitor galaxy and 2. the period in which the global structures are reproduced. 
The SMBH resides close to its apoapsis, implying that the SMBH moves relatively slowly, and the uncertainty in the current position is smaller than in other positions, such as the near periapsis. 

\subsection{Locus of the SMBH}
\label{subsec:hrs:locus}
\begin{figure*}
  \centering
  \epsscale{1.18}
  \plotone{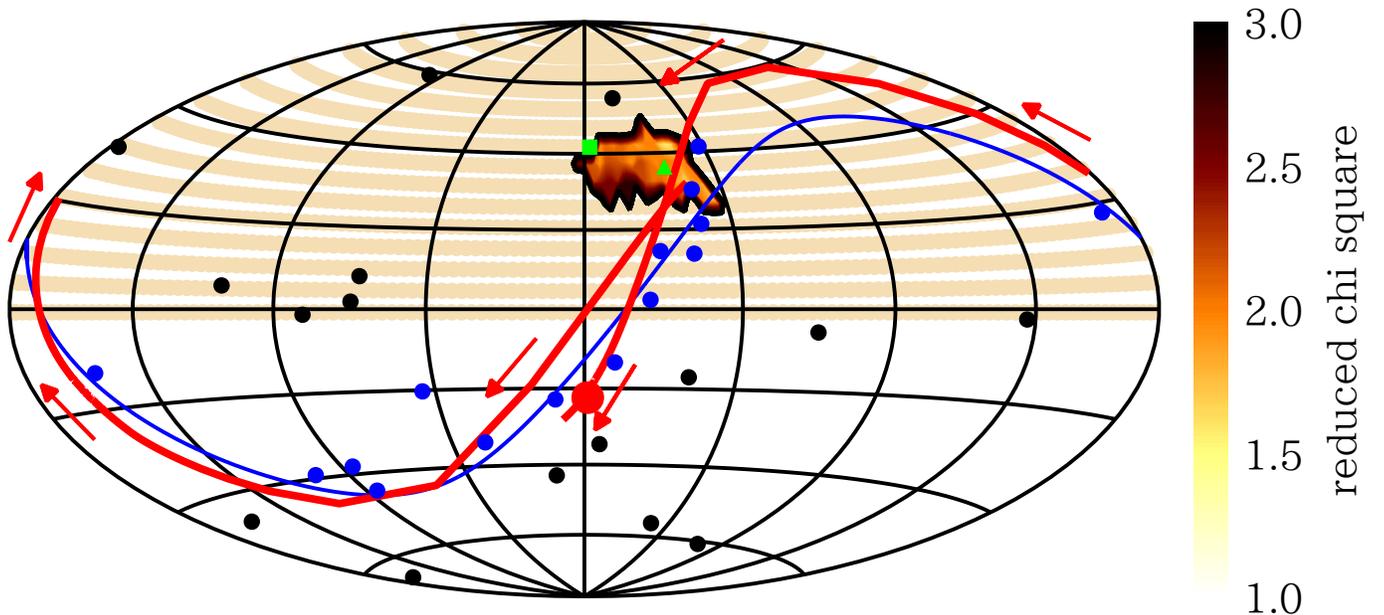}
  \caption{
    The Aitoff-Hammer equal-area projection of satellite galaxy distribution around M31 and infalling orbits of the progenitor. 
    Horizontal and vertical axes indicate the M31-centric galactic longitude and latitude, respectively. 
    Blue circles indicate satellite galaxies distributed in a vast thin disk with a pole at $(l_{\rm M31},\, b_{\rm M31}) = (-78\degr.4,\, 38\degr.3)$ \citep[blue curve:][]{Ibata2013, Conn2013}. 
    Black circles indicate other satellite galaxies listed in \citet{Ibata2013} and \citet{Collins2013}. 
    Light-brown points show infalling satellite orbit models investigated in our low-resolution parameter study; these appear as light-brown bands or zones in the northern hemisphere. 
    Overlaid color map shows the results of the high-resolution parameter study in terms of reduced $\chi^2$ analysis of the shapes of the observed and simulated stellar shells. 
    The green square and triangle show the infalling orbit models of \citet{Fardal2007} and \citet{Sadoun2013}, respectively. 
    All quantities related to the infalling orbit were evaluated at 7.63 kpc from the center of M31 (initial separation of $N$-body simulations in this study). 
    The bold red curve with arrows shows the orbit of an SMBH particle (SMBH ID 3 in Tab.~\ref{tab:mbh}). 
    The SMBH moves along the blue curve (disk plane of the satellites) from the orange-colored region in the bottom left direction. 
    The SMBH progresses along the blue curve and reaches the filled red circle, indicating its current position. \\
    (A three-dimensional view of the SMBH orbits and the distribution of satellite galaxies around M31 is provided in the online-material.)
    }
  \label{fig:survey.hammer}
\end{figure*}

Figure~\ref{fig:survey.hammer} shows the resultant $\chi^2$ map of the 138 runs of the high-resolution $N$-body simulations, together with the spatial distribution of satellite galaxies around M31 in the M31-centric spherical coordinate system defined by \citet{McConnachieIrwin2006}. 
The M31-centric galactic longitude $l_{\rm M31}$ ranges from $-180\degr \leq l_{\rm M31} \leq 180\degr$, where $l_{\rm M31} = 0\degr$ points toward the Milky Way. 
The M31-centric galactic latitude $b_{\rm M31}$ ranges from $-90\degr \leq b_{\rm M31} \leq 90\degr$, where $b_{\rm M31} = 0\degr$ indicates the plane of the M31 disk. 
M31 and the Milky Way locate at the center and at $(l_{\rm M31},\, b_{\rm M31}) = (0\degr,\, -13\degr)$, respectively (not plotted in the figure). 

To compare the infalling orbits tested in this study with the distribution of satellite galaxies, the parameter space of low-resolution $N$-body simulations (light-brown region) and the results of high-resolution $N$-body simulations (color map) are also shown in Fig.~\ref{fig:survey.hammer}. 
The light-brown region covers the northern hemisphere of the M31 halo, while the small area occupied by the color map indicates that the satellite must have infallen within a very narrow directional range in the northern hemisphere to reproduce the observational structures. 
The bold red curve with arrows shows the orbit of an SMBH particle (SMBH ID 3 in Tab.~\ref{tab:mbh}). 
The SMBH moves along the blue curve (disk plane of the satellites) from the orange-colored region in the bottom left direction. 
The SMBH progresses along the blue curve and reaches the filled red circle, indicating its current position. 

We compare our results with those of related studies \citep{Fardal2007, Sadoun2013}. 
For this purpose, Fig.~\ref{fig:survey.hammer} also plots the infalling orbits at $r=7.63$ kpc adopted in earlier studies (\citealt{Fardal2007}; green square; \citealt{Sadoun2013}; green triangle). 
The infalling orbit of \citet{Fardal2007} locates at the edge of the parameter space in the high-resolution $N$-body simulations. 
Earlier, we established that it also locates near the edge of the possible parameter region in terms of the infalling velocity (see Fig.~\ref{fig:survey.result}). 
Thus, this orbit almost matches the orbits of our parameter study. 
Although the infalling orbit of \citet{Sadoun2013} occupies the high-resolution parameter region in Fig.~\ref{fig:survey.hammer}, their orbit is inconsistent with our results. 
This discrepancy arises because the infalling radial velocity is very high, and the orbital angular momentum very low, in their model (as mentioned in \S\ref{subsec:lrs:result}). 

Here, we briefly consider the possibility that the progenitor satellite galaxy entered from the southern hemisphere of M31. 
As mentioned in \S\ref{subsec:lrs:modeling}, to produce the giant stellar stream occupying the southern hemisphere, the progenitor satellite galaxy should fall into the central region of M31 from the northern hemisphere. 
In \S\ref{subsec:lrs:result}, we also demonstrated that the progenitor satellite passed 1 kpc distant from the center of M31 during its free-fall time. 
This indicates that the tidal interaction generated by the bulge of M31 through the pericentric passage is sufficiently strong to destroy the satellite. 
In other words, the merger investigated in this study was the first merger between the satellite and M31. 
Therefore, we conclude that the satellite galaxy entered from the northern hemisphere of M31, as earlier assumed in this study.

\section{Discussion}
\label{section:discussion}
\subsection{Validity of the Assumptions}
\label{subsec:discussion:validity}
\begin{figure}
  \centering
  \epsscale{1.18}
  \plotone{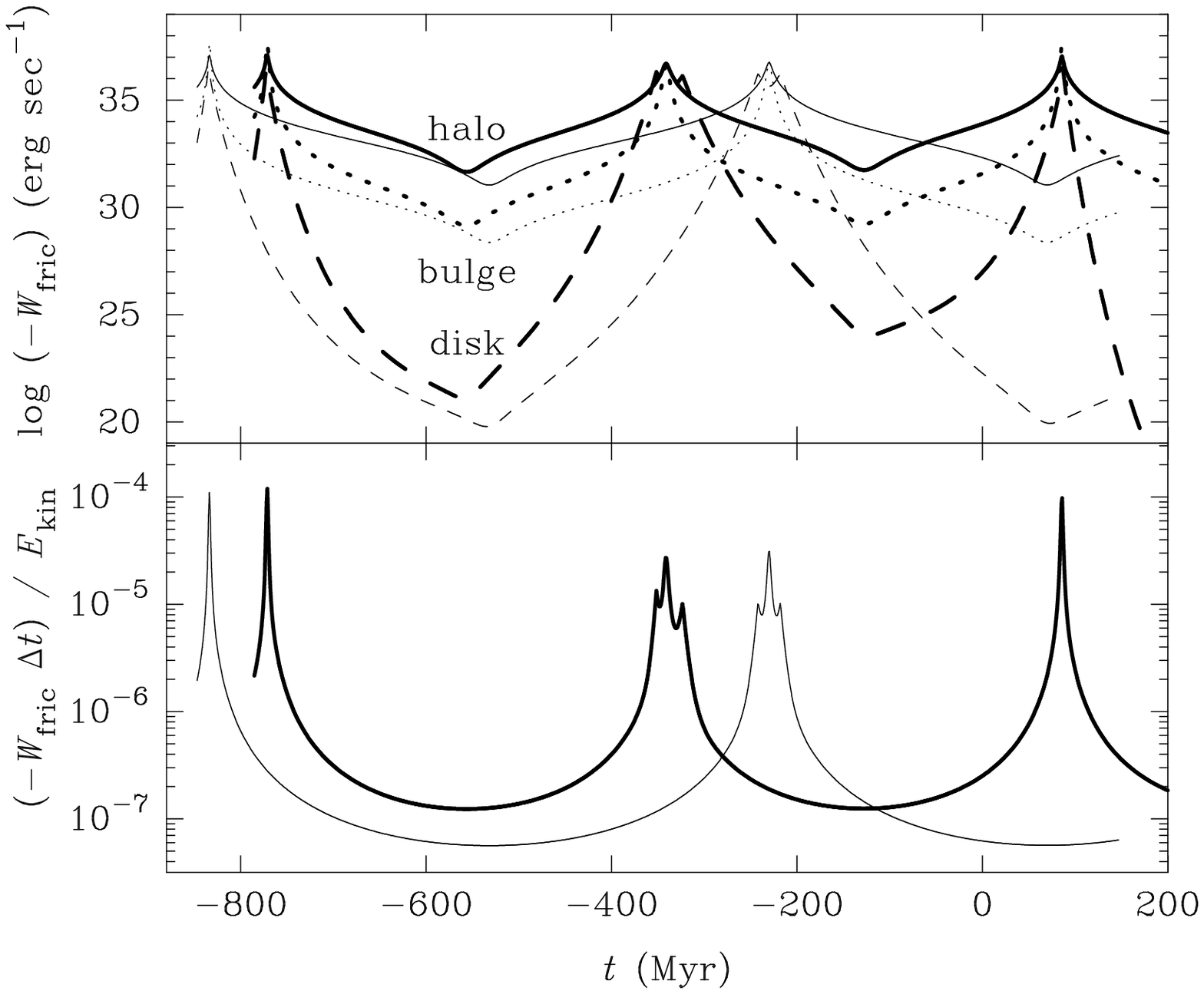}
  \caption{
    Estimating the effects of dynamical friction on the orbital evolution of SMBH particles. 
    In the upper panel, rate at which kinetic energy is lost through dynamical friction is plotted as a function of time. 
    Solid, dotted, and dashed curves show the contributions by M31's halo, bulge, and disk component, respectively. 
    Thin and thick curves show the time evolution of each contribution for two SMBH particles selected from five successful candidates (SMBH ID 1 and 2 in Tab.~\ref{tab:mbh}, respectively). 
    Peaks and troughs in these curves correspond to the passing of SMBHs through their periapsis and apoapsis, respectively. 
    The lower panel plots the energy lost through dynamical friction as a function of time. 
    In these plots, the energy loss during $\Delta t$ of 1 Myr is normalized by the kinetic energy of the particles throughout the same period. 
    Both panels clearly show that the effects of dynamical friction amplify only as the SMBH passes the central region of M31; thus, dynamical friction exerts negligible effect on the motion of SMBHs at the present time. 
    }
  \label{fig:friction}
\end{figure}

In this paper, we assume a fixed potential model for M31. 
Therefore, the simulations exclude the possible effects of dynamical friction introduced by the bulge, the disk, and the halo of the galaxy. 
However, dynamical friction plays a key role in sinking the SMBHs from the halos to the central regions of their host galaxies. 
To estimate how dynamical friction changes the orbits of SMBHs, we now evaluate the amount of kinetic energy lost by the SMBH through dynamical friction. 
The Chandrasekhar formula \citep{BinneyTremaine2008} gives the energy dissipation rate due to dynamical friction $W_{\rm fric}$ as 
\begin{equation}
  W_{\rm fric} = -\frac{4\pi G^2 {M_{\rm BH}}^2 \rho\left(\mbox{\boldmath $r$}_{\rm BH}\right) \ln{\Lambda}}{v_{\rm BH}} \left[{\rm erf}{(X)}-\frac{2X}{\sqrt{\pi}}e^{-X^2}\right].
\end{equation}
Here, $\rho(\mbox{\boldmath $r$})$ is the mass density profile, and $X$ is defined as $v_{\rm BH}/\sqrt{2\hat{\sigma}^2}$, where $\hat{\sigma}$ is the velocity dispersion. The quantities $\ln{\Lambda}$, $\mbox{\boldmath $r$}_{\rm BH}$, and $\mbox{\boldmath $v$}_{\rm BH}$ denote the coulomb logarithm, position, and velocity of the SMBH, respectively. 
To estimate the effects of dynamical friction, we must estimate the mass distribution, the velocity dispersion of field particles, and the coulomb logarithm. 
For this purpose, we adopt the Hernquist bulge, the exponential disk and the NFW halo assumed in the $N$-body simulations. 
The velocity dispersion of the bulge is 260 km s$^{-1}$ \citep{Geehan2006}. 
From the equation of motion and Poisson's equation \citep[see][]{MoriRich2008}, the velocity dispersion of the disk along its rotation axis is obtained as 60 km s$^{-1}$. 
We assumed a thin and axisymmetric disk, with a flat rotation curve and constant velocity dispersion. 
If the distribution function is further assumed as isotropic, the velocity dispersion of the halo is 233 km s$^{-1}$ at its scale radius \citep{Widrow2000}. 
The coulomb logarithm is approximately given by 
\begin{equation}
  \ln{\Lambda} = \ln{\left(\frac{b_{\rm max} {v_{\rm typ}}^2}{GM_{\rm BH}}\right)}, 
\end{equation}
where $b_{\rm max}$ is the maximum impact parameter (here assumed as 40 kpc, sufficiently greater than the size of the bulge and the disk of M31), $v_{\rm typ}$ is the typical velocity (velocity dispersion in the bulge and halo; maximum rotation velocity of 260 km s$^{- 1}$ in the disk). 

Figure~\ref{fig:friction} shows the effects of dynamical friction on the orbital evolution of SMBH particles. 
The rate of energy loss, and the energy loss over 1 Myr (normalized by kinetic energy), evolves as shown in the top and bottom panels, respectively. 
Dynamical friction is relatively large when the SMBH passes its periapsis (a region of high mass density), and exerts negligible effects up to the present day ($t = 0$). 
Since the stream was formed by a near head-on collision (see \S\ref{subsec:lrs:result}), the progenitor resided in the central region of M31 for very short periods. 
This explains why our results were essentially unaffected by dynamical friction. 
As shown in the bottom panel of Fig.~\ref{fig:friction}, about $10^{-3}$ of the SMBH's kinetic energy was dissipated in a single pericentric passage, as the SMBH traversed the central region of M31 over $\sim$10 Myr (see \S\ref{subsec:lrs:result}). 
Thus, dynamical friction encountered during a few crossings exerts no influence on the motion of the SMBH in this study. 
Consequently, our $N$-body simulations, which ignore dynamical friction, are sufficiently realistic to predict the current position of the SMBH. 

In the next place, SMBH shifts from its initial position by gravitational Brownian motion, a random walk in momentum space perturbed by gravitational encounters with nearby stars \citep{Merritt2001, Merritt2005}. 
According to \citet{Merritt2007}, the expected velocity dispersion of the SMBH is 
\begin{equation}
  4.0 \times 10^{-2} \,\mathrm{km\,s}^{-1} \times \left(\frac{\tilde{m}}{1M_\odot}\right)^{1/2} \times \left(\frac{\tilde{\sigma}}{40\,\mathrm{km\,s}^{-1}}\right),
\end{equation}
where $\tilde{m}$ and $\tilde{\sigma}$ are the effective stellar mass and the one-dimensional velocity dispersion of nearby stars in the satellite galaxy, respectively. 
In this paper, we assume that the velocity dispersion of the satellite galaxy is $\tilde{\sigma}=49.1$ km s$^{-1}$ at the center, and $\tilde{\sigma}=39.3$ km s$^{-1}$ at the core radius. 
This value is negligibly small, and will not visibly alter the SMBH orbit; hence, we conclude that excluding these effects does not alter our predictions of the current SMBH position.

\subsection{Origin of the Progenitor Satellite}
\label{subsec:discussion:background}
In the recent observational studies of the satellite galaxy distribution around M31, \citet{Ibata2013} and \citet{Conn2013} concluded that 15 of the satellite galaxies are arranged in a disk-like structure. 
Figure~\ref{fig:survey.hammer} distinguishes the satellite galaxies forming a vast thin disk around M31 \citep{Ibata2013, Conn2013} from other satellite galaxies \citep{Ibata2013, Collins2013}, which are randomly scattered. 
As noted by \citet{Ibata2013} and \citet{Conn2013}, approximately half of the M31 satellite galaxies locate near a disk plane. 
Assuming that M31 was formed by a gas-rich major merger event, and given the observed giant stellar stream \citep{Hammer2010}, \citet{Fouquet2012} and \citet{Hammer2013} proposed that the major merger gave rise to the observed satellite distribution. 
However, the origin of the satellite distribution remains an open question. 

Figure~\ref{fig:survey.hammer} reveals a surprising result: part of the possible orbit region overlaps the disk plane of the satellite galaxies around M31. 
This infers a close connection between the formation process of the giant stellar stream and the disk plane of the satellite galaxies. 
\citet{Sadoun2013} considered that several hundred megayears ago is very recent to admit first-time collision between a bound satellite galaxy and its host galaxy. 
However, this unnatural situation can be understood from the disk-like distribution of satellite galaxies. 
If the satellites are distributed in this way, the probability of satellite-satellite interactions is higher than when the satellites are randomly distributed, since the number density of satellite galaxies is greater in a disk geometry. 
A satellite galaxy within the disk plane can lose most of its angular momentum through inter-satellite interactions. 
Once this occurs, the satellite is expected to follow a highly radial infall orbit toward M31, generating the observed structures such as the giant stellar stream. 
In future studies, we will evaluate the cross-section of these interactions and discuss the possibility of this working hypothesis. 

The remarkable correspondence between the orbit and the disk plane becomes apparent when the orbit of an SMBH particle (SMBH ID 3 in Tab.~\ref{tab:mbh}) is superimposed on the galaxy distribution (see Fig.~\ref{fig:survey.hammer}). 
Two of the five successful orbits listed in Table~\ref{tab:mbh} (SMBH ID 2 and 3) lie on the disk plane. 
Therefore, if future observations establish that the SMBH is wandering in the M31 halo, we will be equipped with highly suggestive clues regarding the formation and evolutionary history of the M31 halo. 
These data will enable connection of the observed stellar structures, and the current and ancient distributions of satellite galaxies. 
Furthermore, since two of the SMBH orbits coincide with the disk plane of the satellite galaxies, more wandering SMBHs might reside in the M31 halo. 
Figure~\ref{fig:survey.hammer} suggests that an SMBH initially moving along the disk plane remains on the plane. 
The observed disk-like satellite distribution is expected to trigger satellite-M31 interactions by extracting orbital angular momentum via unknown process, for example, satellite-satellite interactions. 
Consequently, remnants of ancient merger events should be concentrated in the halo region near the disk plane of the satellites. 
The higher merger rate suggests that SMBHs will be similarly concentrated in this field of the halo. 
Therefore, a group of wandering SMBHs might locate along the disk plane formed by the satellite galaxies. 

\subsection{Impacts on Components of M31}
\label{subsec:discussion:M31components}
Here, the study by \citet{Gordon2006} is worthy of note. 
They concluded that a 10 kpc ring observed in the M31 disk is a remnant of an offset merger. 
The ring structure of radius $\sim$10 kpc has been extensively reported in infrared or H$\alpha$ images \citep{Habing1984, Rice1993, Devereux1994, Haas1998, Gordon2006}. 
\citet{Gordon2006} argued that the ring formed at an offset distance of 1.4 kpc from the center of M31, close to $r_{\rm peri}$ determined in this study (see \S\ref{subsec:lrs:result}). 
This correspondence between the two studies suggests that the 10 kpc ring is another fossil record of the minor merger investigated in this study. 

\citet{Davidge2012} discovered an overdense region of effective radius $0\degr.04$, at $(\xi,\, \eta) = (0\degr.24,\, 0\degr.20)$. 
Our results suggest that the clump found by \citet{Davidge2012} may be related to the giant stellar stream. 
To check the physical connection between the former minor merger event and the clump, the clump must be discriminated from the M31 disk component in phase space, based on spectroscopic observations. 
\citet{Fardal2013} similarly compared the ``location of the progenitor's central material'' with the clump position; however, their results were inconclusive because their stellar particles were widely spaced. 
In Fardal et al.'s study, the particles were initially placed in the central region of the progenitor, and the very central region was resolved by a small number of $N$-body particles, leading to a sparse wider distribution. 
\citet{Kirihara2013?} investigated a model of M31 colliding with a dwarf spiral galaxy comprising a stellar bulge, disk, and dark matter halo. 
They reported that part of the bulge component of the progenitor survived the collision. 
The surviving part is located close to the current position of the wandering SMBH identified in this study. 
This correspondence strongly suggests that the current SMBH position is independent of the morphology of the progenitor satellite galaxy. 

Since the SMBH likely occupies the M31 halo, the SMBH of the progenitor should not be considered as an origin of multiple nuclei in M31* \citep{Lauer1993, Bender2005}. 
Furthermore, since the SMBH locates closer to the Milky Way than the M31 disk, the M31 disk will not disturb future observational attempts to detect the SMBH. 
If the wandering SMBH is experimentally verified in the near future, our understanding of how SMBHs coevolve with their host galaxies will be greatly enhanced. 
The candidate fields determined in this study will complement future observations. 
Since the mass of the SMBH is close to the low-mass end of the $M_{\rm BH}-\sigma$ relation, observational discovery of the SMBH will provide information on the low-mass end of the SMBH-host galaxy associations. 

\section{Conclusion}
\label{section:conclusion}
We investigated an SMBH in an ancient satellite galaxy, whose current position is consistent with the observed structures in the M31 halo. 
The infalling orbit of the satellite was first established by conducting numerous low-resolution parameter trials on a high performance GPU cluster. 
These preliminary investigations reduced the possible parameter space for the orbit to a manageable size. 
Next, the orbital evolution of the SMBH was directly calculated in high-resolution $N$-body simulations. 
The hermitage of the SMBH was localized to the northeast stellar shell over an area of $\sim 0\degr.6 \times 0\degr.7$. 
The observational field $1\degr \times 1\degr$ was sufficiently wide to contain all possible positions of the wandering SMBH. 
Furthermore, we found signatures of the relationships between this particular minor merger and the recently identified thin disk plane formed by M31's satellite galaxies. 
This discovery may assist in identifying a group of wandering SMBHs in the halo of M31. 
Our forthcoming paper will present a feasibility study on detecting the wandering SMBHs \citet{Kawaguchi2013?}. 

\acknowledgments

We thank A. Tanikawa for useful suggestion about $N$-body simulations. 
Numerical Simulations have been performed with HA-PACS at the Center for Computational Sciences, University of Tsukuba. 
This work was partially supported by the program of the Pre-Strategic Initiatives, University of Tsukuba, the FIRST project based on Grants-in-Aid for Specially Promoted Research by MEXT (16002003), and the Grant-in-Aid for Scientific Research (S)(20224002), (A)(21244013), (C)(18540242), and (C)(25400222).

\appendix
\section{Caption of online-only material}
The caption of the three-dimensional view of Figure \ref{fig:survey.hammer} is as follows.\\
A three-dimensional view of the SMBH orbits and the distribution of satellite galaxies around M31 is provided in the online-material (Back perspective corresponds to a view from LSR). 
Three-dimensional visualization was conducted with the S2PLOT programming library \citep{Barnes2006, BarnesFluke2008}.
Colored spheres with bars, vectors and labels indicate satellite galaxies distributed around M31 with errors of distance measurements, heliocentric line-of-sight velocity and object name: red spheres denote satellites forming the thin disk, blue spheres denote other satellites listed in \citet{Ibata2013} and yellow spheres indicate satellites listed in \citet{Collins2013} as well as M32 and NGC 205. 
Cyan and green curves show the 138 candidate orbits: cyan curves result from the $N$-body simulations; green curves are orbits of the SMBH falling from the corresponding apoapsis into $r = 7.63$ kpc (evaluated by test-particle calculations). 
Magenta and orange curves highlight the five successful orbits and dark gray spheres show the current position of SMBHs. 
A movie version of the three-dimensional view of Figure \ref{fig:survey.hammer} is also provided as an online-material. 




\begin{thebibliography}{}
\bibitem[Barnes et al.(2006)]{Barnes2006} Barnes, D.~G., Fluke, C.~J., Bourke, P.~D., \& Parry, O.~T.\ 2006, \pasa, 23, 82 
\bibitem[Barnes \& Fluke(2008)]{BarnesFluke2008} Barnes, D.~G., \& Fluke, C.~J.\ 2008, \na, 13, 599 
\bibitem[Barth et al.(2005)]{Barth2005} Barth, A.~J., Greene, J.~E., \& Ho, L.~C.\ 2005, \apjl, 619, L151
\bibitem[Bellovary et al.(2010)]{Bellovary2010} Bellovary, J.~M., Governato, F., Quinn, T.~R., et al.\ 2010, \apjl, 721, L148 
\bibitem[Bender et al.(2005)]{Bender2005} Bender, R., et al.\ 2005, \apj, 631, 280 
\bibitem[Binney \& Tremaine(2008)]{BinneyTremaine2008} Binney, J., 
  \& Tremaine, S.\ 2008, Galactic Dynamics: Second Edition, 
  by James Binney and Scott Tremaine, ~ISBN 978-0-691-13026-2 (HB).
  ~Published by Princeton University Press, Princeton, NJ USA, 2008
\bibitem[Bullock \& Johnston(2005)]{BullockJohnston2005}Bullock, J. S., \& Johnston, K. V. 2005, \apj, 635, 931
\bibitem[Chiba et al.(2005)]{Chiba2005} Chiba, M., Minezaki, T., Kashikawa, N., Kataza, H., \& Inoue, K.~T.\ 2005, \apj, 627, 53 
\bibitem[Collins et al.(2013)]{Collins2013} Collins, M.~L.~M., Chapman, S.~C., Rich, R.~M., et al.\ 2013, \apj, 768, 172 
\bibitem[Conn et al.(2013)]{Conn2013} Conn, A.~R., Lewis, G.~F., Ibata, R.~A., et al.\ 2013, \apj, 766, 120 
\bibitem[Davidge(2012)]{Davidge2012} Davidge, T.~J.\ 2012, \apjl, 749, L7 
\bibitem[de Vaucouleurs et al.(1991)]{deVaucouleurs1991} de Vaucouleurs, 
G., de Vaucouleurs, A., Corwin, H.~G., Jr., et al.\ 1991, Third Reference 
Catalogue of Bright Galaxies.
\bibitem[Devereux et al.(1994)]{Devereux1994} Devereux, N.~A., Price, R., Wells, L.~A., \& Duric, N.\ 1994, \aj, 108, 1667 
\bibitem[Falc{\'o}n-Barroso et al.(2011)]{FalconBarroso2011} Falc{\'o}n-Barroso, J., van de Ven, G., Peletier, R.~F., et al.\ 2011, \mnras, 417, 1787 
\bibitem[Fardal et al.(2006)]{Fardal2006} Fardal, M.~A., Babul, A., Geehan, J.~J., \& Guhathakurta, P.\ 2006, \mnras, 366, 1012 
\bibitem[Fardal et al.(2007)]{Fardal2007} Fardal, M.~A., Guhathakurta, P., Babul, A., \& McConnachie, A.~W.\ 2007, \mnras, 380, 15
\bibitem[Fardal et al.(2012)]{Fardal2012} Fardal, M.~A., Guhathakurta, P., Gilbert, K.~M., et al.\ 2012, \mnras, 423, 3134 
\bibitem[Fardal et al.(2013)]{Fardal2013} Fardal, M.~A., Weinberg, M.~D., Babul, A., et al.\ 2013, \mnras, 434, 2779 
\bibitem[Farrell et al.(2009)]{Farrell2009} Farrell, S.~A., Webb, N.~A., Barret, D., Godet, O., \& Rodrigues, J.~M.\ 2009, \nat, 460, 73 
\bibitem[Ferguson et al.(2002)]{Ferguson2002} Ferguson, A.~M.~N., Irwin, M.~J., Ibata, R.~A., Lewis, G.~F., \& Tanvir, N.~R.\ 2002, \aj, 124, 1452 
\bibitem[Font et al.(2006)]{Font2006} Font, A.~S., Johnston, K.~V., Guhathakurta, P., Majewski, S.~R., \& Rich, R.~M.\ 2006, \aj, 131, 1436 
\bibitem[Fouquet et al.(2012)]{Fouquet2012} Fouquet, S., Hammer, F., Yang, Y., Puech, M., \& Flores, H.\ 2012, \mnras, 427, 1769 
\bibitem[Geehan et al.(2006)]{Geehan2006} Geehan, J.~J., Fardal, M.~A., Babul, A., \& Guhathakurta, P.\ 2006, \mnras, 366, 996 
\bibitem[Gilbert et al.(2007)]{Gilbert2007} Gilbert, K.~M., et al. 2007, \apj, 668, 245 
\bibitem[Gilbert et al.(2009)]{Gilbert2009} Gilbert, K.~M., et al. 2009, \apj, 705, 1275 
\bibitem[Gordon et al.(2006)]{Gordon2006} Gordon, K.~D., Bailin, J., Engelbracht, C.~W., et al.\ 2006, \apjl, 638, L87 
\bibitem[Guhathakurta et al.(2006)]{Guhathakurta2006} Guhathakurta, P., et al. 2006, \aj, 131, 2497 
\bibitem[Haas et al.(1998)]{Haas1998} Haas, M., Lemke, D., Stickel, M., et al.\ 1998, \aap, 338, L33 
\bibitem[Habing et al.(1984)]{Habing1984} Habing, H.~J., Miley, G., Young, E., et al.\ 1984, \apjl, 278, L59 
\bibitem[Hammer et al.(2010)]{Hammer2010} Hammer, F., Yang, Y.~B., Wang, J.~L., et al.\ 2010, \apj, 725, 542 
\bibitem[Hammer et al.(2013)]{Hammer2013} Hammer, F., Yang, Y., Fouquet, S., et al.\ 2013, \mnras, 431, 3543 
\bibitem[Hernquist(1990)]{Hernquist1990} Hernquist, L. 1990, \apj, 356, 359 
\bibitem[Ibata et al.(2001)]{Ibata2001} Ibata, R., Irwin, M., Lewis, G., Ferguson, A.~M.~N., \& Tanvir, N.\ 2001, \nat, 412, 49 
\bibitem[Ibata et al.(2004)]{Ibata2004} Ibata, R., Chapman, S., Ferguson, A.~M.~N., Irwin, M., Lewis, G., \& McConnachie, A.\ 2004, \mnras, 351, 117 
\bibitem[Ibata et al.(2005)]{Ibata2005} Ibata, R., Chapman, S., Ferguson, A.~M.~N., Lewis, G., Irwin, M., \& Tanvir, N.\ 2005, \apj, 634, 287 
\bibitem[Ibata et al.(2007)]{Ibata2007} Ibata, R., Martin, N.~F., Irwin, M., Chapman, S., Ferguson, A.~M.~N., Lewis, G.~F., \& McConnachie, A.~W.\ 2007, \apj, 671, 1591 
\bibitem[Ibata et al.(2013)]{Ibata2013} Ibata, R.~A., Lewis, G.~F., Conn, A.~R., et al.\ 2013, \nat, 493, 62 
\bibitem[Irwin et al.(2005)]{Irwin2005} Irwin, M.~J., Ferguson, A.~M.~N., Ibata, R.~A., Lewis, G.~F., \& Tanvir, N.~R.\ 2005, \apjl, 628, L105 
\bibitem[Kalirai et al.(2006a)]{Kalirai2006a} Kalirai, J.~S., Guhathakurta, P., Gilbert, K.~M., Reitzel, D.~B., Majewski, S.~R., Rich, R.~M., \& Cooper, M.~C.\ 2006, \apj, 641, 268 
\bibitem[Kalirai et al.(2006b)]{Kalirai2006b} Kalirai, J.~S., et al. 2006, \apj, 648, 389 
\bibitem[Kalirai et al.(2010)]{Kalirai2010} Kalirai, J.~S., Beaton, R.~L., Geha, M.~C., et al.\ 2010, \apj, 711, 671 
\bibitem[Kawaguchi et al.(in prep.)]{Kawaguchi2013?} Kawaguchi, T. et al. in prep.
\bibitem[Kirihara et al.(in prep.)]{Kirihara2013?} Kirihara, T., Miki, Y., \& Mori, M. in prep.
\bibitem[Koch et al.(2008)]{Koch2008}Koch et al. 2008, \apj, 689, 958
\bibitem[Lauer et al.(1993)]{Lauer1993} Lauer, T.~R., et al. 1993, \aj, 106, 1436
\bibitem[Magorrian et al.(1998)]{Magorrian1998}Magorrian et al. 1998, \aj, 115, 2285
\bibitem[Marconi \& Hunt(2003)]{MarconiHunt2003}Marconi, A., \& Hunt, L. K. 2003, \apj, 589, L21
\bibitem[Martin et al.(2013)]{Martin2013} Martin, N.~F., Ibata, R.~A., McConnachie, A.~W., et al.\ 2013, \apj, 776, 80 
\bibitem[McConnachie et al.(2003)]{McConnachie2003} McConnachie, A.~W., Irwin, M.~J., Ibata, R.~A., Ferguson, A.~M.~N., Lewis, G.~F., \& Tanvir, N.\ 2003, \mnras, 343, 1335 
\bibitem[McConnachie \& Irwin(2006)]{McConnachieIrwin2006} McConnachie, A.~W., \& Irwin, M.~J.\ 2006, \mnras, 365, 902 
\bibitem[McConnachie et al.(2009)]{McConnachie2009} McConnachie, A.~W., et al.\ 2009, \nat, 461, 66
\bibitem[Merritt(2001)]{Merritt2001} Merritt, D.\ 2001, \apj, 556, 245 
\bibitem[Merritt(2005)]{Merritt2005} Merritt, D.\ 2005, \apj, 628, 673 
\bibitem[Merritt et al.(2007)]{Merritt2007} Merritt, D., Berczik, P., \& Laun, F.\ 2007, \aj, 133, 553 
\bibitem[Minezaki et al.(2009)]{Minezaki2009} Minezaki, T., Chiba, M., Kashikawa, N., Inoue, K.~T., \& Kataza, H.\ 2009, \apj, 697, 610 
\bibitem[Miki et al.(in prep.)]{Miki2013?} Miki, Y., Mori, M., \& Rich, R. M. in prep.
\bibitem[Miki et al.(2013)]{Miki2013}Miki, Y., Takahashi, D., \& Mori, M.\ 2013, Computer Physics Communications, 184, 2159
\bibitem[Mori \& Rich(2008)]{MoriRich2008}Mori, M., \& Rich, R. M. 2008, \apj, 674, L77
\bibitem[Navarro et al.(1996)]{NFW1996} Navarro, J.~F., Frenk, C.~S., \& White, S.~D.~M.\ 1996, \apj, 462, 563 
\bibitem[Press et al.(2007)]{Press2007} Press, W. H., Teukolsky, S. A., Vetterling, W. T., \& Flannery, B. P.\ 2007, Numerical Recipes: Third Edition, ~ISBN 978-0-521-88407-5.
\bibitem[Rice(1993)]{Rice1993} Rice, W.\ 1993, \aj, 105, 67
\bibitem[Richardson et al.(2011)]{Richardson2011} Richardson, J.~C., Irwin, M.~J., McConnachie, A.~W., et al.\ 2011, \apj, 732, 76 
\bibitem[Sadoun et al.(2013)]{Sadoun2013} Sadoun, R., Mohayaee, R., \& Colin, J.\ 2013, arXiv:1307.5044 
\bibitem[Sohn et al.(2012)]{Sohn2012} Sohn, S.~T., Anderson, J., \& van der Marel, R.~P.\ 2012, \apj, 753, 7 
\bibitem[Tanaka et al.(2010)]{Tanaka2010}Tanaka, M., Chiba, M., Komiyama, Y., Guhathakurta, P., Kalirai, J. S., \& Iye, M. 2010, \apj, 708, 1168
\bibitem[Tollerud et al.(2012)]{Tollerud2012} Tollerud, E.~J., Beaton, R.~L., Geha, M.~C., et al.\ 2012, \apj, 752, 45 
\bibitem[Toloba et al.(2012)]{Toloba2012} Toloba, E., Boselli, A., Peletier, R.~F., et al.\ 2012, \aap, 548, A78 
\bibitem[van der Marel et al.(2012)]{vanderMarel2012} van der Marel, R.~P., Fardal, M., Besla, G., et al.\ 2012, \apj, 753, 8 
\bibitem[Widrow(2000)]{Widrow2000} Widrow, L.~M.\ 2000, \apjs, 131, 39 
\bibitem[Wiersema et al.(2010)]{Wiersema2010} Wiersema, K., Farrell, S.~A., Webb, N.~A., Servillat, M., Maccarone, T.~J., Barret, D., \& Godet, O.\ 2010, \apjl, 721, L102
\bibitem[Xiao et al.(2011)]{Xiao2011} Xiao, T., Barth, A.~J., Greene, J.~E., et al.\ 2011, \apj, 739, 28 
\end{thebibliography}
\end{document}